\documentclass[twocolumn,prl, aps,superscriptaddress,showpacs]{revtex4-2}
\usepackage{mathptmx}
\DeclareMathAlphabet{\mathcal}{OMS}{cmsy}{m}{n}
\usepackage{subfigure}
\usepackage{psfrag}
\usepackage{dcolumn}
\usepackage{amsmath,amssymb}
\usepackage{bm}
\usepackage{color}
\usepackage{latexsym}
\usepackage{epstopdf}
\usepackage{color}
\usepackage[english]{babel}
\usepackage{latexsym}

\usepackage{graphicx}
\usepackage{epsf}
\usepackage{subfigure}
\usepackage{amsmath}
\usepackage{amssymb}
\usepackage{amsfonts}
\usepackage{bm}
\usepackage{natbib}
\usepackage{epstopdf}
\usepackage{appendix}
\usepackage{lipsum, babel}
\usepackage{soul}

\definecolor{mygrey}{gray}{0.35}
\definecolor{myblue}{rgb}{0.2,0.2,0.8}
\definecolor{myzard}{cmyk}{0,0,0.05,0}
\definecolor{mywhite}{rgb}{1,1,1}
\definecolor{myred}{rgb}{1,0.,0.3}
\definecolor{myblack}{rgb}{0,0,0}

\usepackage{hyperref}
\hypersetup{colorlinks=true,citecolor=myblue,linkcolor=blue,urlcolor=myblue}

\def\be{\begin{equation}}
\def\ee{\end{equation}}
\def\ba{\begin{align}}
\def\enda{\end{align}}
\def\bi{\begin{itemize}}
\def\ei{\end{itemize}}

 \def\ee{\mathord{\rm e}}

\def\beq{\begin{equation}}
\def\eeq{\end{equation}}
\def \bml{\begin{multline}}
\def \eml{\end{multline}}
\def \bea{\begin{eqnarray}}
\def \eea{\end{eqnarray}}

\newcommand{\sens}{|S\rangle}

\newcommand{\kp}{k_{\mathrm{s}}}
\newcommand{\kc}{k_\mathrm{c}}

\newcommand{\vT}{v_\mathrm{T}}

\newcommand{\up}{\mid\uparrow\rangle}
\newcommand{\down}{\mid\downarrow\rangle}

\newcommand{\downb}{\langle\downarrow\mid}

 \newcommand{\ket}[1]{|#1\rangle}

\newcommand{\bla}[1]{\left(#1\right)}
\newcommand{\blb}[1]{\left[#1\right]}

\newcommand{\bld}[1]{\left\{#1\right\}}

\newcommand{\kd}{k_{\mathrm{d}}}

\DeclareMathOperator{\sinc}{sinc}
\usepackage{dsfont}

\begin{document}

\widetext


\title{Continuous protection of a collective state from inhomogeneous dephasing  }
\author{R. Finkelstein}
\author{O. Lahad}
\affiliation{Physics of Complex Systems, Weizmann Institute of Science, Rehovot 7610001, Israel}
\author{I. Cohen}
\affiliation{Department of Physics and Astronomy, Aarhus University, Ny Munkegade 120, DK-8000 Aarhus C, Denmark}
\author{O. Davidson}
\author{S. Kiriati}
\author{E. Poem}
\author{O. Firstenberg}
\affiliation{Physics of Complex Systems, Weizmann Institute of Science, Rehovot 7610001, Israel}

\begin{abstract}
We introduce and demonstrate a scheme for eliminating the inhomogeneous dephasing of a collective quantum state. The scheme employs off-resonant fields that continuously dress the collective state with an auxiliary sensor state, which has an enhanced and opposite sensitivity to the same source of inhomogeneity. We derive the optimal conditions under which the dressed state is fully protected from dephasing, when using either one or two dressing fields. The latter provides better protection, circumvents qubit phase rotation, and suppresses the sensitivity to drive noise. We further derive expressions for all residual, higher-order, sensitivities. We experimentally study the scheme by protecting a collective excitation of an atomic ensemble, where inhomogeneous dephasing originates from thermal motion. Using photon storage and retrieval, we demonstrate complete suppression of inhomogeneous dephasing and consequently a prolonged memory time. Our scheme may be applied to eliminate motional dephasing in other systems, improving the performance of quantum gates and memories with neutral atoms. It is also generally applicable to various gas, solid, and engineered systems, where sensitivity to variations in time, space, or other domains limits possible scale-up of the system.

\end{abstract}

\maketitle



The quantum state of a system is prone to decoherence via inhomogeneous dephasing due to variations among the system's constituents.
These variations include spatial inhomogeneities, primarily nonuniform external fields \cite{Lepoutre2018b,Munowitz1986,Kotler2011}, environmental imperfections \cite{Pingault2014,Cogan2018}, finite temperature effects such as a thermal velocity distribution \cite{finkelstein2019,finkelstein2018,Kaplan2002,Levine2018,Saffman2016}, and fabrication infidelities in engineered systems \cite{Senellart2017,Krantz2019,Devoret2013}. Slow temporal fluctuations and shot-to-shot variations, also relevant to single-constituent systems, may as well manifest as inhomogeneous dephasing  \cite{Levine2018,Manovitz2019,Aharon2013,Bermudez2012}. In quantum information processing, these variations limit the qubit coherence time.

In some cases, such as in dual-color magic-wavelength optical traps, it is possible to minimize inhomogeneous dephasing by introducing an additional field that induces the exact same inhomogeneity and thus directly balances the differential phase shifts \cite{Harber2002,Ye2008,Radnaev2010,Lahad2019,Hilton2019}.
In more general cases, inhomogeneous dephasing can be mitigated by pulse-based protocols, namely echo sequences and various dynamical decoupling methods~\cite{Viol1999,Khodjasteh2005,Souza2011,Rui2015}. In these protocols, the system regains coherence only at discrete, pre-determined times. Several protocols of continuous dynamical decoupling have been studied~ \cite{Fanchini2007,Tan2013,Bermudez2012,Golter2014}, typically relying on strong resonant driving of the qubit transition. Such driving is not compatible with all systems and requires additional measures for eliminating the sensitivity to drive noise.  

Here we propose and demonstrate an alternative approach for mitigating inhomogeneous dephasing. Our scheme employs off-resonance fields for continuously dressing the target state with an additional \emph{sensor} state, which has an opposite and enhanced sensitivity to the same source of inhomogeneity. This admixture of a portion of the sensor state forms a protected, dephasing-free state.
Notably, the dressing field is a continuous wave, providing continuous protection and not limiting the extraction of quantum information to predefined times.


Experimentally, we focus on optical excitations in thermal atomic ensembles. Storing and processing quantum information as collective excitations in atomic ensembles 
offers long coherence times \cite{Zhao2009,Dudin2013,Yang2016,Korber2018,Katz2018}, noiseless memory protocols \cite{finkelstein2018,kaczmarek2018}, remote entanglement \cite{Julsgaard2001,Krauter2013,Lettner2011}, and the generation of non-classical light \cite{Duan2001,Boyer2008}. Nevertheless, whether they are warm or ultracold, the atoms are not stationary; their inhomogeneous velocity distribution leads to dephasing of collective states formed by excitations with non-zero momentum transfer \cite{Whiting2017,Saffman2016,Levine2018}. Such motional dephasing is a prevailing decoherence mechanism, yet it is rarely mitigated by means beyond actual cooling and trapping of the atoms. While some pulsed spin-echo techniques have been proposed \cite{Moiseev2001} and demonstrated \cite{Rui2015}, a robust and continuous protocol for protection from motional dephasing has not been realized to date.

In the following, we first present a general analysis of the protection scheme with either one or two dressing fields. We identify the minimal requirements and conditions for optimal protection. We show that introducing two dressing fields reduces not only the sensitivity to frequency variations, but also to fluctuations in the dressing fields, known as drive noise \cite{Tan2013,Stark2018,Spielman2018}. We then turn to the case study of motional dephasing in atomic ensembles. Our experimental realization is based on light storage in a fast ladder memory (FLAME) \cite{finkelstein2018}, using the retrieval efficiency as a measure of atomic coherence~\cite{Phillips2001}. We demonstrate complete cancellation of motional dephasing, prolonging the memory lifetime, and verify the scaling of optimal conditions for protection. We further confirm that the double-dressing scheme, in contrast to the single-dressing scheme, inflicts no qubit phase rotation and is thus less sensitive to drive noise, providing robust continuous protection.

\begin{figure}[ht!]
    \centering
    \includegraphics[width=\columnwidth]{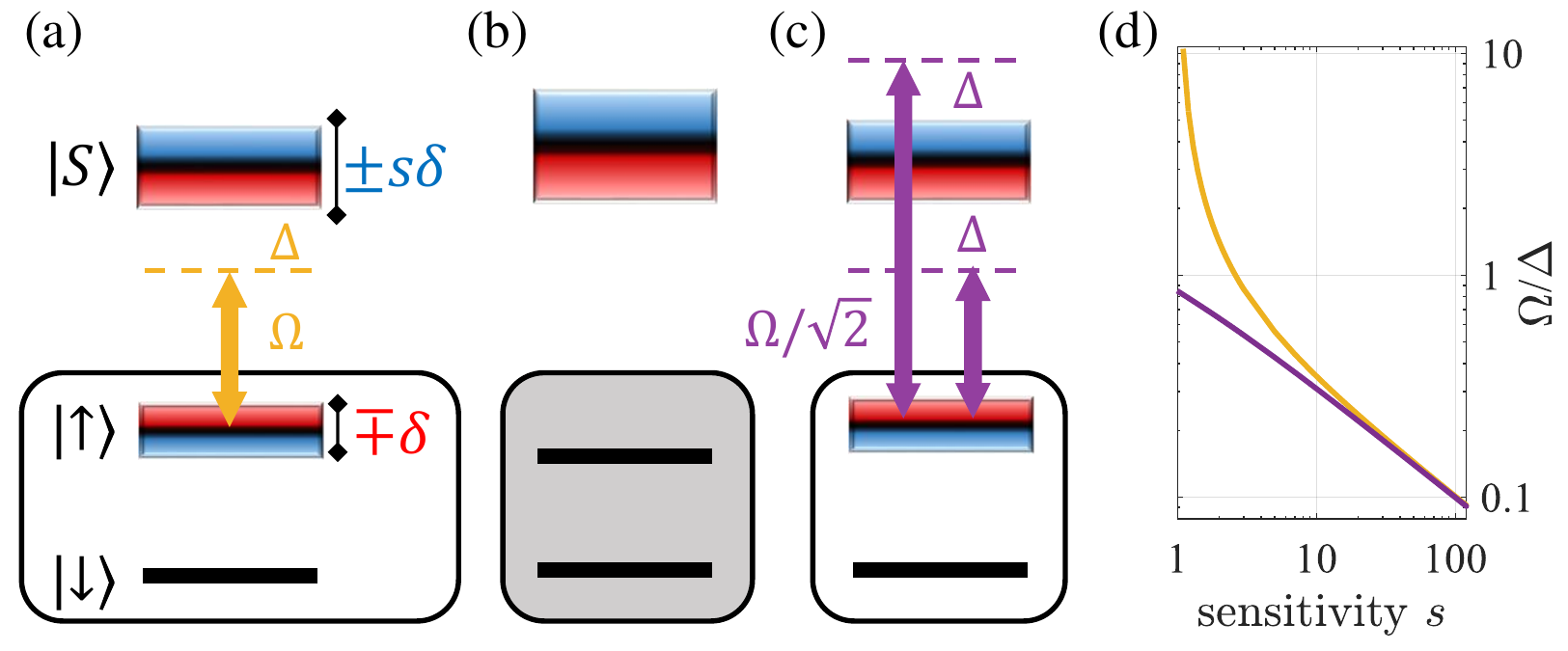}
    \caption{\textbf{Continuous protection of a quantum state.} (a) Without protection, the transition frequency of the qubit $\down-\up$ is shifted by $\delta$ due to some source of inhomogeneity, while an auxiliary sensor state $\sens$ experiences an opposite and possibly larger shift $-s\delta$. A single field with Rabi frequency $\Omega$ and detuning $\Delta$ dresses the qubit with the sensor state. (b) A protected qubit is formed under conditions of optimal dressing. The transition frequency of the protected qubit is slightly altered by the mean light shift. (c) Applying two dressing fields of equal intensities and opposite detunings further eliminates the mean shift and reduces the sensitivity of the protected qubit to dressing noise. (d) Optimal dressing ratio $\Omega/\Delta$ as a function of the sensitivity parameter $s$ for the single dressing (orange) and double dressing (purple) schemes. }
    \label{fig:Intro}
\end{figure} 
%
\subsection*{Continuous protection using a sensor state}
Consider a qubit comprising the states $\down$ and $\up$, whose transition frequency experiences an inhomogeneous shift $\delta$ [Fig.~\ref{fig:Intro}(a)]. We require that one of these states, here chosen to be $\up$ without loss of generality, can be coupled by an external dressing field to a third state $\sens$, which acts as a sensor. We further require that the overall transition frequency between $\down$ and $\sens$ be sensitive to the same source of inhomogeneity, such that it experiences an opposite and possibly larger shift, $-s\delta$, where $s$ is the sensitivity factor. For example, in the case of motional dephasing, $s$ is the ratio between the Doppler widths of the transitions $\down-\sens$ and $\down-\up$, and, in the case of magnetically sensitive transitions, $s$ is the ratio between their magnetic moment differences. 
If the above requirements are satisfied, the dressing field can be tuned to protect the qubit from inhomogeneous dephasing.

We begin by considering a single dressing field with Rabi frequency $\Omega$ and detuning $\Delta$ [Fig.~\ref{fig:Intro}(b)]. The states $\up$ and $\sens$ are mixed, forming dressed states. With respect to $\down$, the states $\sens$ and $\up$ are inhomogeneously shifted by  \mbox{$[\Delta+\delta(s-1)]/2\pm\sqrt{[\Delta+(s+1)\delta]^2/4 +\Omega^2}$}. Optimal protection is achieved when the transition frequency of the qubit ($\down-\up$) is insensitive to variations in $\delta$ to first order, which occurs when
\begin{equation}
\label{single-dress cond}
\Omega^2/\Delta^2=s/(s-1)^2.    
\end{equation}
This condition is shown in Fig.~\ref{fig:Intro}(d) by the orange line. Since the ratio $\Omega^2/\Delta^2$ determines the magnitude of mixing between the bare states, the larger the sensitivity factor $s$ is, the smaller the portion of $\sens$ admixed into $\up$ at optimal protection.

For $s\gg1$, the dressed state frequency-shift takes the form of a light shift:  $\Omega^2/(\Delta+s\delta)\approx (\Omega^2/\Delta)[1-s\delta/\Delta+O(s\delta/\Delta)^2]$. It is the first-order term $(\Omega/\Delta)^2s\delta$ which counteracts the inhomogeneous shift $\delta$ when condition \eqref{single-dress cond} is fulfilled. Given a distribution of inhomogeneous shifts $\delta$ of width $\sigma$, a dressing field detuning $\Delta\gg s\sigma$ allows efficient protection for practically all $\delta$. This regime has been recently used for narrowing and enhancement of spectral lines \cite{Lahad2019,finkelstein2019}.

A caveat of the single-dressing scheme is the mean, zeroth-order shift $\Omega^2/\Delta$ added to the transition frequency of the protected qubit. This shift rotates the qubit phase in time, exposing it to noise in the drive field $\Omega$. To avoid this, two dressing fields at opposite detunings $\pm\Delta$ should be used [Fig.~\ref{fig:Intro}(c)]. The mean shifts they induce are opposite in sign and thus cancel. In fact, as all even-order terms cancel, a protection up to third order in $\delta/\Delta$ is achieved.

The above arguments carry on to any $s$. For a general solution with more than one dressing field, we use the Magnus expansion of the time-dependent Hamiltonian and find the condition under which the transition frequency is insensitive to variations in $\delta$ to first and second order:
\begin{equation}
\label{double-dress cond}
    J_0(2\sqrt{2}\Omega/\Delta)=(s-1)/(s+1),
\end{equation}
where $J_0$ is the zeroth-order Bessel function (see Supplementary Information for the full derivation). Equation \eqref{double-dress cond} converges to Eq.~\eqref{single-dress cond} for $s\gg 1$, but $\Omega/\Delta$ scales more favorably with $s$ for small $s$, such that less Rabi frequency and less mixing are required with double dressing [see Fig.~\ref{fig:Intro}(d)].

\begin{figure}[t!]
\centering
\includegraphics[width=\columnwidth]{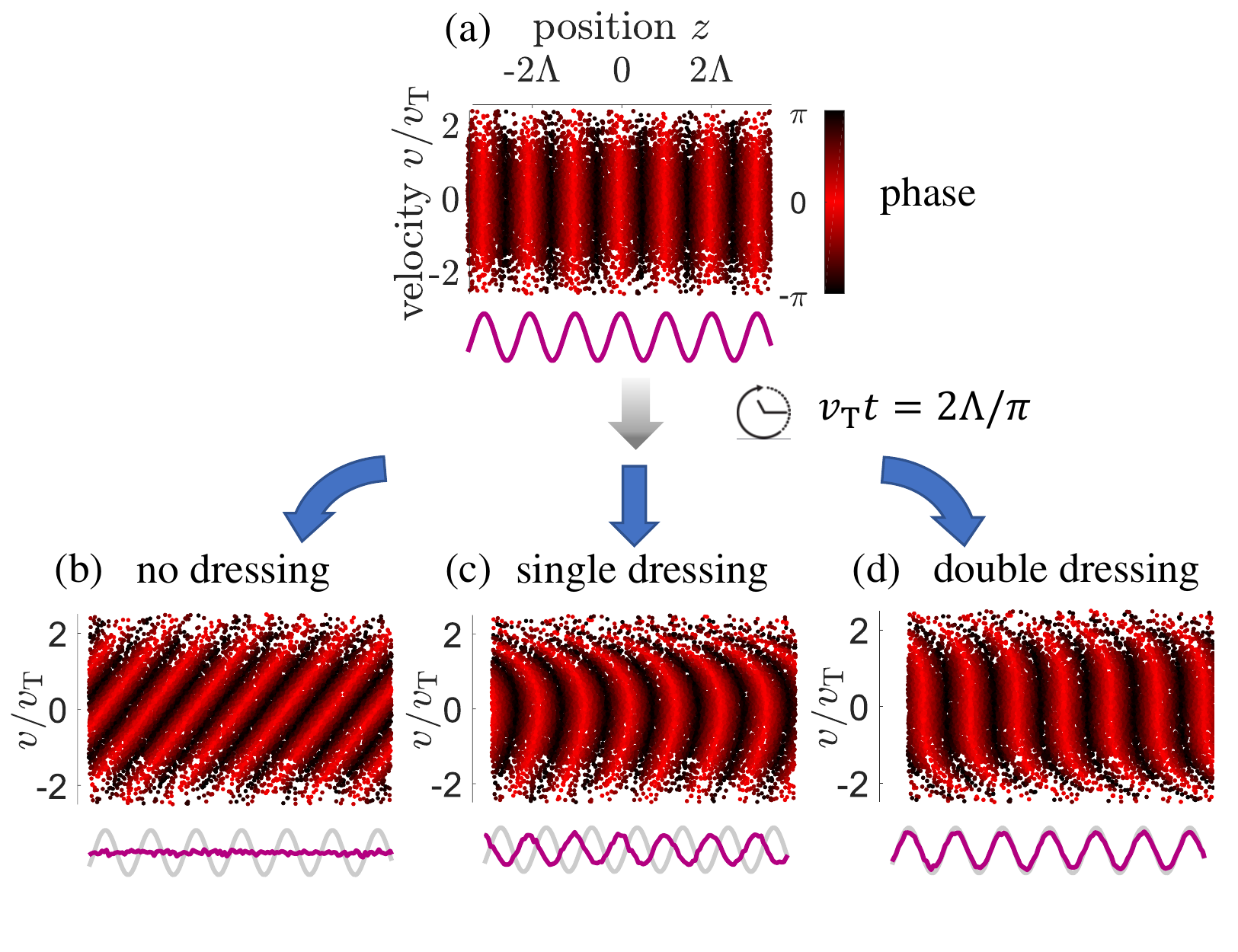} 
\caption{\textbf{Motional dephasing and protection of a spin wave.} (a) Position-velocity representation of a collective excitation (spin wave), generated, {\it{e.g.}}, by light storage in a two-photon bi-chromatic transition. The collective state has a spatially dependent phase with a wavelength $\Lambda$ (top), clearly apparent when the real part of the spin-wave is averaged over all velocities (purple, bottom). (b) Thermal atomic motion (mean velocity $\vT$) at time $t=2\Lambda/(\pi\vT)$ after the initial excitation. Averaging over velocities results in decoherence of the collective state. (c) In the presence of a single dressing field, atoms with different velocities experience different light shifts, and the spatial coherence is maintained up to second order in velocity. The zeroth-order term introduces a global phase shift, evident when comparing the spin wave (purple) to its original form (gray). (d) When two dressing fields are applied, the protection is up to third order in velocity (thus covering a wider velocity range), and no global phase shift is introduced. Calculations are performed with $s=110$, $\Delta=4v_T/\Lambda$, and optimal $\Omega$ from Eqs.~(1) and (2).
}
\label{fig:dephasing}
\end{figure}
\subsection*{Motional dephasing}
Initializing an ensemble of atoms in a collective state is typically done by optical excitation. When the wavevector $k=2\pi/\Lambda$ of the collective excitation is nonzero, a spatial phase $e^{i(2\pi /\Lambda )z}$ is imprinted on the atoms, as illustrated in Fig.~\ref{fig:dephasing}(a). In multi-photon transitions, the excitation wavevector is the vectorial sum of the participating fields' wavevectors. For a single excitation, the collective state can be written as a spin wave $|W\rangle=\frac{1}{\sqrt{N}}\sum_j{e^{i(2\pi /\Lambda )z_j} \up_j\downb_j |G\rangle}$, where $z_j$ is the position of atom $j$ at storage time $t=0$, and $|G\rangle=\prod_j\down_j$ is the initial collective ground state.
When the atoms are displaced by thermal motion, atoms with different thermal velocities $v_j$ carry the original phase to different positions $z_j+v_jt$ along the spin wave [Fig.~\ref{fig:dephasing}(b)]. The resulting state is
\begin{equation}\label{Dephasing_eqn}
    |W'(t)\rangle=\sum_j{e^{i(2\pi /\Lambda )(z_j-v_j t)} \up_j \downb_j |G\rangle},
\end{equation}
whose overlap with $|W\rangle$, quantified by the squared coherence $\mathcal C(t)\equiv |\langle W|W'(t)\rangle|^2$, determines the retrieval efficiency of light into the phase-matched direction $\eta(t)=\eta(0)\mathcal{C}(t)$ \cite{Whiting2017}. Here $\eta(0)$ is the combined efficiency of the read and write processes. For pure motional dephasing, $\mathcal C(t)=e^{-(t/T_{\mathrm{inhom}})^2}$, where  $T_{\mathrm{inhom}}=\sqrt{2}/\sigma=\Lambda/(\sqrt{2}\pi\vT)$ is the inhomogeneous dephasing time, and $\vT$ is the atomic thermal velocity.

To apply the continuous protection scheme in the case of motional dephasing, we dress the spin-wave with a sensor state which has large and opposite sensitivity to velocity. This is achieved by an optical transition whose wavevector is larger than $2\pi/\Lambda$. The frequency shift of the transition $\down-\up$ due to the dressing field depends on the atom velocity via the Doppler effect and, for optimal dressing, exactly cancels the bare motional dephasing, rendering a velocity-insensitive state. Remarkably, as shown in Figs.~\ref{fig:dephasing}(c,d), although the atoms are constantly in motion during the evolution of the state, the spin-wave correlations between position and phase are largely maintained. This unique time-evolution is animated in Supplementary Video 1.

Apart from protecting against dephasing of the spin wave (when $|W\rangle\rightarrow |W'\rangle$), as quantified by the light retrieval efficiency, we are interested in protecting a general qubit $\alpha |G\rangle + \beta |W\rangle$. This qubit is formed, for instance, by mapping a photonic qubit $\alpha |0\rangle + \beta |1\rangle$, where $|0\rangle$ and $|1\rangle$ are photon number states, onto atomic collective states. Decoherence of $|W\rangle$ would naturally lead to decoherence of the qubit, but so will global fluctuations of the relative phase of $\beta/\alpha$ \cite{Schmidt-Eberle2019}. The double dressing scheme, which does not introduce a global phase that vary with drive noise, is therefore advantageous for robust protection of a qubit. 

\begin{figure}[t!] \includegraphics[width=\columnwidth]{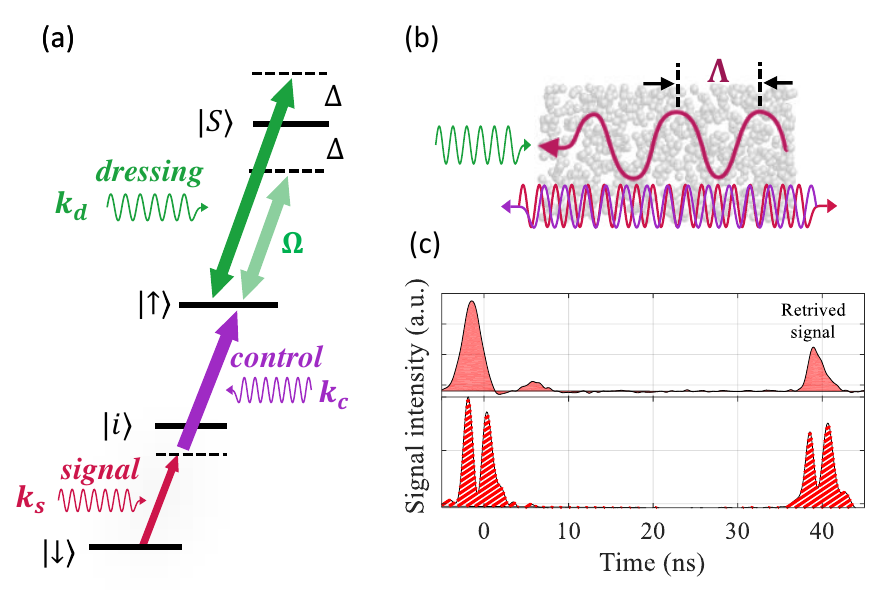}
\caption{\textbf{Experimental realization of continuous protection from motional dephasing.} (a) Light storage scheme: Counter-propagating pulses of a strong control field and a weak signal field with wavevectors $\kc$  and $\kp$ couple the ground state $\down$ to the excited state $\up$ through an intermediate state $|i\rangle$. A dressing beam with wavevector $\kd$ couples $\up$ to the sensor state $\sens$ with Rabi frequency $\Omega$ and detuning $\Delta$. (b) The signal and control pulses generate a collective excitation with wavevector $2\pi/\Lambda = \kp-\kc$. The dressing field wavevector $\kd$ has an opposite direction and a larger magnitude. (c) Typical data traces when the incoming and retrieved signals are either directly measured (top) or interfered with reference pulses to measure the phase of the signal (bottom).\label{fig:Exp_setup}}
\end{figure}

\subsection*{Experimental realization} To demonstrate and study the continuous protection against motional dephasing, we probe the coherence decay of a collective excitation in a hot atomic vapor, utilizing storage and retrieval of light. We employ the FLAME protocol \cite{finkelstein2018} to store 1.8-ns (FWHM) signal pulses (with $\sim 0.5$ photons per pulse) as electronic-orbital excitations  [Fig.~\ref{fig:Exp_setup}(a)] in a 5-mm-long cell of thermal $^{87}\mathrm{Rb}$ vapor at 98$^{\circ}$C. The $\up$ and $\down$ states reside in the levels $5S_{1/2}$ and $5D_{5/2}$ respectively. The two states are coupled via a two-photon transition, detuned by 1.4~GHz from the intermediate level $5P_{3/2}$ [$|i\rangle$ in Fig.~\ref{fig:Exp_setup}(a)]. After some storage time $t$, we send a second control pulse and retrieve a signal pulse from the collective excitation [Fig. 3(c) top]. The memory efficiency $\eta(t)/\eta(0)$ is determined from the retrieved power (relative to the input power), and the temporal decay of the squared coherence is calculated from $\mathcal{C}(t)=\eta(t)/\eta(0)$. To measure the phase difference between incoming and retrieved signals, we store slightly longer signal pulses (4-ns FWHM) and interfere them with reference pulses shifted by $2\pi\cdot 380$ MHz, which do not traverse the atomic vapor [Fig.~\ref{fig:Exp_setup}(c) bottom]. 

The wavevectors for the signal and control transitions are respectively $\kp=2\pi/0.78~\mathrm{\mu m}^{-1}$ and $\kc=2\pi/0.776~\mathrm{\mu m}^{-1}$, giving an excitation wavelength $\Lambda=2\pi/(\kp-\kc)= 151~\mathrm{\mu m}$ [Fig.~\ref{fig:Exp_setup}(b)]. The Gaussian distribution of velocities with a thermal velocity $v_T=188$~m/s results in an inhomogeneous dephasing of the collective state over about 100 ns. Protection during storage is obtained by a field with wavevector $\kd=2\pi/1.274~\mathrm{\mu m}^{-1}$, propagating along the direction of the signal field and weakly dressing the state $\up$ with the state $\sens=28F_{7/2}$. We use an electro-optic modulator (EOM) to generate the double dressing configuration (see Methods). The sensor transition $\down-\sens$ has an opposite and enhanced velocity-sensitivity $s=(\kp-\kc+\kd)/(\kp-\kc) \approx 110$ compared to that of the two-photon transition $\down-\up$, such that weak dressing is sufficient to form a velocity-insensitive collective state.
The bare homogeneous decoherence rate of the collective state $\gamma=2\pi\cdot 1.38$ MHz is limited by the $5D_{5/2}$ radiative lifetime and by the transit time of atoms through the signal beam waist. In the presence of a dressing beam, we measure an increase of $\le 25$\% in $\gamma$, which we attribute to resonant scattering due to imperfect extinction of the carrier frequency by the EOM and to residual excitation of atoms in the tail of the velocity distribution for small dressing field detunings (see Methods).

\begin{figure}[tb!]
    \centering
    \includegraphics[width=\columnwidth]{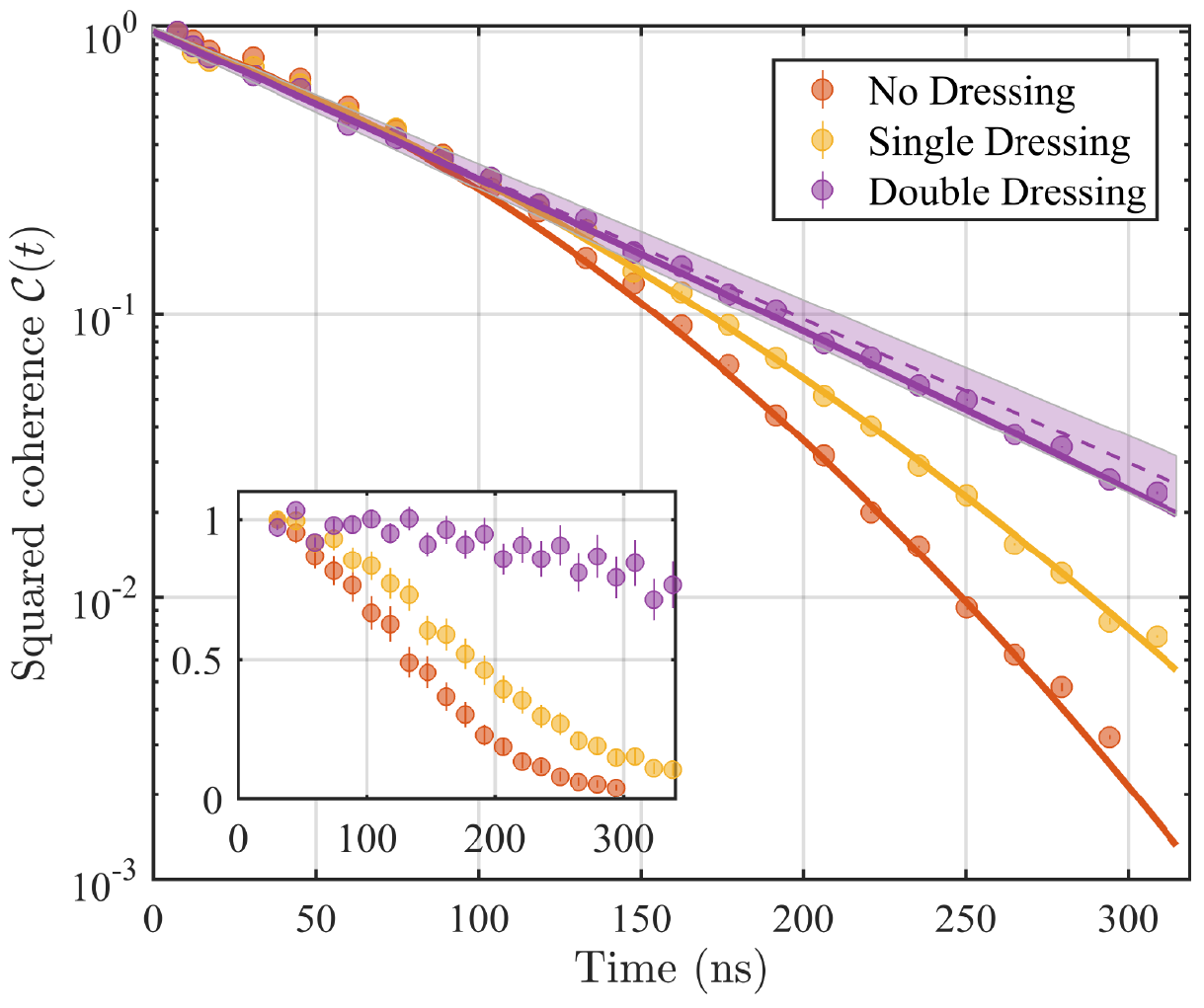}
    \caption{\textbf{Protection of a collective excitation from inhomogeneous dephasing.} The squared coherence is calculated for different storage times by measuring the memory efficiency. Solid lines are fit to a model accounting for both homogeneous and inhomogeneous dephasing. With double dressing, the decay is predominantly homogeneous; the dashed line marks the homogeneous component of the fitted decay model, with surrounding shaded area representing fit uncertainty. Inset: The inhomogeneous component of the decoherence $\mathcal{C}(t)/e^{-\gamma t}$, where the homogeneous component $e^{-\gamma t}$ is normalized out. The dressing field parameters are $|\Delta|=2\pi\cdot350$ MHz, $\Omega=2\pi\cdot34$ MHz (double dressing), and $\Omega=2\pi\cdot28$ MHz (single dressing). \label{fig:Time}}
  
\end{figure}

To quantify the effect of protection, we compare the decay of coherence with no dressing field, with a single dressing field, and with a double dressing field; these are shown in Fig.~\ref{fig:Time}. In the absence of dressing fields, the decay of efficiency with time has both exponential and Gaussian components, corresponding respectively to homogeneous and inhomogeneous dephasing. With a single dressing field, the inhomogeneous component is partially removed, while with the double dressing field, it vanishes almost completely. As a result, we observe an increase of over an order of magnitude in retrieved signal at long storage times. 

The residual inhomogeneous dephasing is revealed by normalizing the squared coherence $\mathcal{C}(t)$ by the homogeneous decoherence $e^{-\gamma t}$, as plotted in the inset of Fig.~\ref{fig:Time}. With a double-dressing protection field, the inhomogeneous dephasing time is substantially elongated. Importantly, this implies that a dramatic improvement in the coherence times can be obtained for the broad range of systems that enjoy low homogeneous decoherence rate. The remaining, non-linear dependence of the qubit transition frequency on the inhomogeneous shift $\delta$ results in non-Gaussian dephasing \cite{Sung2019}. On long time scales, the coherence decays as a power-law $\mathcal C(t) \propto t^{-2/n}$, where $n$ is the first non-vanishing order of $\delta$ in the qubit transition frequency. However on a short time scale, the dephasing is well approximated by a Gaussian decay (see Supplementary Information). 

 \begin{figure}[t!] 
\centering
\includegraphics[width=\columnwidth]{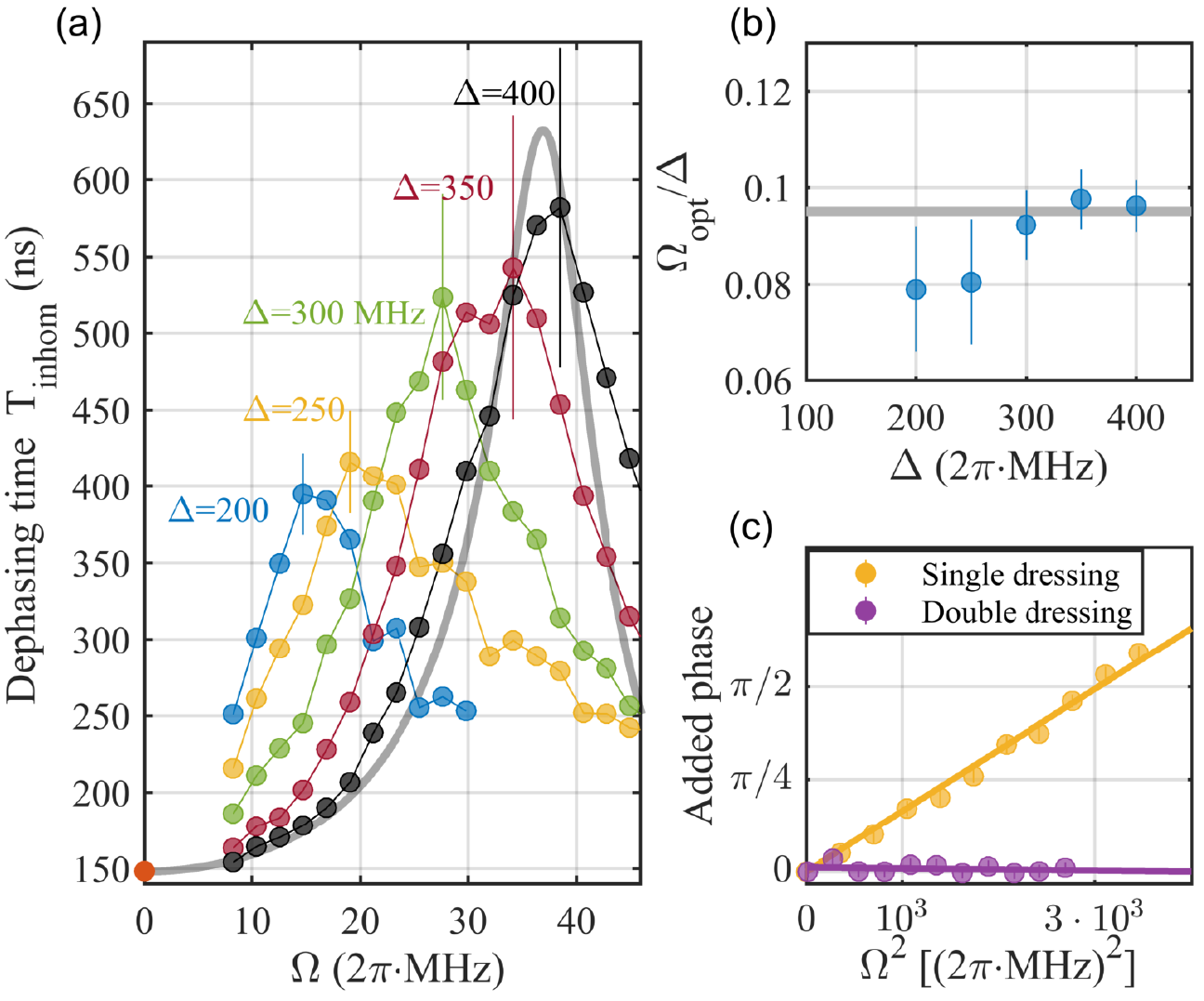}
\caption{\textbf{Dependency of coherence time and accumulated phase on the dressing parameters.} (a) Inhomogeneous dephasing time $T_{\mathrm{inhom}}$, with double-dressing protection as a function of dressing Rabi frequency $\Omega$, for different detunings $\Delta$ (in $2\pi\cdot$MHz). Shaded gray line is a theoretical model, where the dephasing rates due to linear and cubic dependencies on $\delta$ are summed in quadrature. (b) The ratio $\Omega_{\mathrm{opt}}/\Delta$, where $\Omega_{\mathrm{opt}}$ is the Rabi frequency maximizing the gain for each $\Delta$, compared to the model prediction $2^{-3/2}J_0^{-1}[(s-1)/(s+1)]\approx s^{-1}\approx 0.095$ (solid line). (c) Phase accumulated due to the dressing fields during a storage time of 40 ns. Here $\Delta=2\pi\cdot480$ MHz. Solid lines are linear fits. For double dressing, the added phase is zero (up to the experimental uncertainty) and independent of the dressing intensity.}
\label{fig:DressPower}
\end{figure}

To further study the optimal conditions for continuous protection, we vary the dressing parameters $\Omega$ and $\Delta$ and measure the retrieval efficiency at $t=250$ ns, from which we infer the dephasing time $T_{\mathrm{inhom}}$ (see Methods). Figure \ref{fig:DressPower}(a) shows the improvement in $T_{\mathrm{inhom}}$ for the double-dressing protection scheme. The data agrees with the condition $|\Delta|>s\sigma$ for efficient protection (here $s\sigma=2\pi\cdot140$ MHz). Substantial improvement in dephasing time is obtained over a large range of dressing powers and is further enhanced with larger detuning. For the largest detuning measured, we compare the data to the theoretical model (shaded gray line). Indeed, the improvement in $T_{\mathrm{inhom}}$ is quadratic in $\Omega$ and reaches an optimal value at the protection condition \eqref{double-dress cond}, where it is limited by the third-order term in the inhomogeneous shift. The latter can be made negligible by increasing the detuning. We further validate the protection condition \eqref{double-dress cond} in Fig.~\ref{fig:DressPower}(b) by comparing the predicted ratio $\Omega/\Delta$ to the measured ratio $\Omega_{\mathrm{opt}}(\Delta)/\Delta$. Excellent agreement is found for large detunings.

An important property of any protection or dynamical decoupling technique is the qubit phase introduced by the driving fields and its dependence on their intensities. Measurements of this dependence for both single and double dressing are shown in Fig.~\ref{fig:DressPower}(c). For single dressing, we find indeed that the phase shift scales linearly with the mean light shift $\Omega^2/\Delta$ \cite{Parniak2019}, yielding a sensitivity to drive noise of $\sim$ 0.4 mrad/$(2\pi\cdot$MHz$)^2$. This is in contrast to the double dressing protection, where we measure essentially no added phase and no dependence on the dressing intensity or on the storage time.

\subsection*{Potential application to other systems}
Motional dephasing occurs across various platforms, where it limits the performance of quantum information applications. One prominent example is that of neutral atoms trapped in tweezer arrays and interacting via Rydberg excitations. The coherence time and bell-state preparation fidelity in several recent realizations of this system are limited by motional dephasing \cite{Levine2018,Graham2019}. Our scheme can be adapted to mitigate this effect by using a uniform field for globally dressing either the ground or the Rydberg state with an auxiliary meta-stable state or by implementing velocity-insensitive bichromatic entangling gates. An entirely different application is a single photon source based on four-wave mixing in atomic vapor. Whether they rely on Rydberg interactions or not \cite{Ripka2018,Lee2016}, the brightness and purity of these  sources are limited as well by motional dephasing, which can be suppressed by implementing our scheme. Here, when it comes to stochastic events such as the heralded spontaneous generation of photons, a protection scheme that is continuous is essential. For collective Rydberg excitations, a three-photon excitation to the Rydberg state can be employed with an intermediate (two-photon) transition that is velocity sensitive (\emph{i.e.}, with larger sensitivity than in our experiment); this seemingly peculiar choice would yield a state which is protected from motional dephasing and, at the same time, has a large Rydberg admixing, such that it inherits the desired Rydberg-level qualities, namely strong dipolar interactions.

Beyond the context of motional dephasing, our scheme can be applied to magnetically sensitive qubits, such as trapped ions and NV centers. Here again, one should identify a sensor transition with opposite and possibly larger magnetic moment difference relative to the qubit transition. For example, we suggest protecting an optical qubit in a $^{88}\mathrm{Sr}^+ $ ion encoded in the $\ket{S_{1/2};m_j{=}1/2}-\ket{D_{5/2};m_j{=}1/2}$ basis by dressing the ground state with the long-lived $\ket{D_{5/2};m_j{=}-3/2}$ state. This choice provides a sensitivity factor $s=6$ owing to the different land\'{e} $g$-factors of the states involved and thus entails low admixing of the sensor level into the qubit state.

%
\subsection*{Discussion} The presented protection scheme is akin to continuous dynamical decoupling but distinct in several aspects. Our protecting fields do not couple the two states comprising the qubit, but rather couple one of them to an external state. They are preferably off-resonant, and the overall protection improves with further detuning. In standard continuous dynamical decoupling, on the other hand, the protecting fields drive the qubit transition, often resonantly, exposing it directly to drive noise $\delta_\Omega$ when the field amplitude fluctuates $\Omega \rightarrow \Omega+\delta_\Omega$. Moreover, if $\delta$ fluctuates in time, the fluctuation bandwidth over which the protection is effective is limited by the available Rabi frequency \cite{Stark2017}. In our scheme, the bandwidth is fundamentally limited by the detuning (or  the effective Rabi frequency $\sqrt{4\Omega^2+\Delta^2}$), which can be made much larger given the same limited power, especially with large sensitivity $s$ (for which $\Delta \gg \Omega$) 

In Table \ref{tab:summary}, we present the effectiveness of our protection schemes in terms of the residual (higher-order) sensitivities to inhomogeneous shift O$(\delta^2)$,O$(\delta^3)$; to Rabi frequency fluctuations O$(\delta_\Omega)$; and to cross-terms of the two O$(\delta\delta_\Omega)$. For a single dressing field, an enhanced sensitivity $s\gg1$ earned from utilizing a sensor state enables the use of a protection field that is far-detuned, thus reducing sensitivity to drive noise to $\delta_\Omega/\sqrt{s}$. Further suppression of drive noise is achieved with double dressing, which we find to be insensitive to drive noise to leading order O($\delta_\Omega$) [also supported by the measurements in Fig. \ref{fig:DressPower}(c)]. Our analysis additionally shows that the double dressing removes sensitivity to inhomogeneous shifts up to O$(\delta^3)$. The remaining sensitivity scales inversely with Rabi frequency or with detuning, but we note that it scales unfavourably with the sensitivity factor $s$. The dephasing time $T_{\mathrm{inhom}}$ is proportional to the inverse of the leading residual sensitivity term when substituting $\delta=\sigma$.  

For a sensor state with sensitivity equal to that of the qubit state, $s=1$, the single-dressing protection condition [Eq. \eqref{single-dress cond}] requires a resonant field $\Delta=0$, which results in two, equally mixed, protected states \cite{Stark2018,Spielman2018}. The double dressing scheme also generates two protected states in this case (see Supplementary Information), without the need for resonant driving. The $s=1$ case resembles the case of a two-level system, where the condition \eqref{double-dress cond} provides for intrinsic dynamical decoupling in bichromatic entangling gates with trapped ions \cite{Sutherland2019}.  

\begin{table}[tb!]
\caption{\textbf{Leading residual sensitivities of the qubit transition frequency:} first-order in the drive noise ($\delta_\Omega$), second- and third- order in the inhomogeneous shift ($\delta^2$, $\delta^3$), and the cross term (${\delta_\Omega\cdot\delta}$). For the double dressing scheme, exact numerical factors depend on the experimental realization, including the phase between the two fields. These factors and their exact derivation are given in the Supplementary Information.  }\label{tab:summary}
\begin{tabular}{ |c||c|c||c|c|c|c|  }
\colrule
 & \multicolumn{2}{c||}{{single dressing}} & \multicolumn{4}{c|}{{double dressing}} \\ [0.5ex]

& O($\delta_\Omega$) &  O($\delta^2$) & O($\delta_\Omega$) &  O($\delta^2$) & O($\delta \delta_\Omega$) &  O($\delta^3$) \\ [0.5ex]
\colrule
& & & & & &\\ [-2ex]
 $s=1$	& $\delta_\Omega$ & $\frac{1}{2}\frac{\delta^2}{\Omega}$
 & 0 & 0 & $\sim1\frac{\delta\delta_\Omega}{\Omega}$ & $\sim\frac{1}{2}\frac{\delta^3}{\Omega^2}$\\[1ex]

 $s\gg 1$ &	$\frac{2}{\sqrt{s}}\delta_\Omega$ & $\sqrt{s}\frac{\delta^2}{\Omega}$	
 & 0	& 0 & $2\frac{\delta\delta_\Omega}{\Omega}$ & $\sim s\frac{\delta^3}{\Omega^2}$\\[1ex]
\colrule
 \end{tabular}
 \end{table}
 

In conclusion, we have introduced and demonstrated a new scheme for protection from inhomogeneous dephasing, which is continuous, efficient, and robust to drive noise. The scheme is particularly suitable for protection from motional dephasing of a collective state stored in an atomic ensemble, where limited solutions were suggested to date. The minimal requirements outlined in this paper may be found across numerous systems where a multi-level structure exists. We have discussed a few examples for such systems, where this scheme can be applied to enhance the performance of quantum sensors, sources and gates. 

Our experimental demonstration confirms the validity of the scheme by eliminating the inhomogeneous dephasing of a collective excitation in a gas of thermal atoms. Remarkably, this is achieved not through time-reversal of the process of dephasing or of the direction of atomic motion, but through continuously maintaining the original position-dependent phases of atoms which move randomly to different positions.  

The compatibility of this scheme to various protocols of quantum information processing, including gates and metrology, requires further research. It could provide an important tool for these applications operating with either single-constituent qubits or with collective quantum states of ensembles. 

\subsection*{Methods} \emph{Experimental design. --}
The setup comprises a 780 nm distributed Bragg reflector (DBR) diode laser, serving as the signal beam, and a 776 nm external cavity diode laser (ECDL) amplified by a tapered amplifier (TA), serving as the control beam. The signal laser is offset-locked to a master ultra-stable fiber laser using a fast beat-note detector. The control laser is locked to a two-photon absorption feature in a reference cell when overlapped with the master laser, where the latter is frequency shifted by a fiber electro-optic phase modulator (EOPM). The signal field is amplitude modulated in time by two fiber electro-optic amplitude modulators (EOAM) to carve a Gaussian pulse of 1.8 ns FWHM, with a combined extinction ratio of 1:3000. The control field is amplitude modulated by two Pockels cells (PCs), generating pulses of 2.5 ns FWHM with an extinction ratio of 1:1000. The repetition rate of the experiment is set by that of the PCs to 100~kHz. After the modulators, the control beam is passed through a tilted filter  (Semrock LL01-780-12.5) functioning as a 776 nm bandpass, filtering out other frequencies that might be produced in the TA. Both beams are passed through single mode fibers (SMF), aligned with each other in a counter-propagating geometry, and overlapped at the center the vapor cell. The signal beam is focused down to $w_0=85~\mu$m, while the control beam is focused down to $w_0=200~\mu$m. Both beams are $\sigma^+$ polarized, and we optically pump all atoms to the stretched state, a combination which guarantees purely orbital transitions and greatly simplifies the multi-level structure of the atomic vapor. The optical pumping is realized by a `pump' and a `repump' at 795 nm, both $\sigma^+$ polarized, and separated from  each other by 6.8 GHz, such that the pump (repump) is resonant with the $F=2 \rightarrow F'=2$ ($F=1 \rightarrow F'=2$) transition of the rubidium D1 line.  The pump (repump) beam power at the vapor cell is 300~mW (200~mW). The pump beams are 1.2 mm wide and are directed at a small angle with respect to the control beam.
The 5-mm-long $^{87}$Rb vapor cell is anti-reflection coated for 780-1064 nm. It is heated to 72$^\circ$C at its coldest spot and 98$^\circ$C at its hottest spot using two electrical current heaters, to set a Rb density of $6.5\times10^{11}$~cm$^{-3}$ and an optical depth OD$\approx5$. We obtain OD$\approx8.5$ with continuous optical pumping.
After the cell, the signal beam is passed through a polarizing beam-splitter and two 780 nm bandpass filters to filter out any residual 776 nm and 795 nm components. It is then coupled to a SMF acting as a spatial filter, removing most of the spatially incoherent fluorescence emitted from the cell at 780 nm. The SMF is coupled either to a fast linear avalanche photo detector (APD) with bandwidth of 1~GHz, or to a single photon counting module (SPCM) connected to a time tagger with time bins of 100~ps.

\emph{Dressing fields generation.--} The dressing beam is produced by a 1274 nm ECDL and is then passed through both an EOAM and an EOPM. For the double-dressing protection scheme, the EOPM is modulated by a square-wave electronic signal (70 ps rise/fall time), generating dominant first-order side bands comprising 78\% of the total output power and a vanishingly small  carrier ($0.2\%$). The dressing beam is then amplified by an O-band booster optical amplifier (Thorlabs BOA1130S) and further amplified by a TA. The dressing laser frequency is stabilized to a wavelength meter with 1 MHz resolution (High-Finesse WS8-IR1). The dressing beam is combined with the signal beam on a dichroic mirror and focused to $w_0=200~\mu$m in the vapor cell. It is also $\sigma^+$ polarized, thus coupling only to the stretched state.

\emph{Homogeneous decoherence in the experiment.--}
With a single dressing field and, more significantly, with a double dressing field, we measure a small increase in the homogeneous decoherence rate (of up to 25\%). The first contribution to this increase is the scattering rate of atoms at the tale of the inhomogeneous distribution. For the intermediate detunings we use in the experiment $|\Delta| \lesssim 3\sigma$, there is a small but non-vanishing fraction of atoms whose transition frequency is near-resonant with the dressing field. The overall scattering rate due to this contribution is on the order of $1.2{-}6.3\cdot10^6 ~\mathrm{s}^{-1}$ in the range of parameters used in the experiment and scales inversely with $\Delta^2$. 

For the double dressing scheme, the experimental realization (as detailed above) includes a single laser tuned to resonance and a fast square-wave phase modulation, whose imperfections result in a carrier intensity of up to $\Omega^2_\mathrm{carrier}/\Omega^2=0.3(2)\%$. This corresponds to a calculated resonant scattering rate of up to $0.6\cdot 10^5~ \mathrm{s}^{-1}$ (at the optimal Rabi frequency of $\Omega=2\pi\cdot 35$ MHz). 

\emph{Evaluation of $T_{\mathrm{inhom}}$ .--}
To obtain the inhomogeneous dephasing time, we measure the memory efficiency for a long storage time of 250 ns. The results of this measurement are presented in the Supplementary Information (Fig. S2). We extract $T_{\mathrm{inhom}}$ from the efficiency by fitting to a Gaussian decay model $e^{-(t/T_{\mathrm{inhom}})^2}$ after normalizing out the homogeneous decoherence $e^{-\gamma t}$. We include the slight increase in $\gamma$ in the presence of the dressing field by measuring $\gamma$ for several dressing powers at $\Delta=2\pi\cdot350$ MHz (as in Fig. 4 in the main text) and calibrate our model for resonant scattering $\gamma=\gamma_0+\Gamma\Omega^2/\Delta^2$ accordingly. We verify the validity of the model across a wide range of dressing field parameters as presented in Fig. S2 (Supplementary Information). Larger available laser power and faster modulation of the dressing field would enable operation at larger detunings which would greatly suppress such scattering.  

\emph{Phase measurements.--} In order to measure the phase between the incoming and retrieved signals, we generate two consecutive signal pulses and pass them through an acousto-optical modulator (AOM). The diffracted pulses, shifted by $2\pi\cdot380$ MHz relative to the signal, will act as a reference. They are thus coupled into an optical fiber acting as a delay line. The time difference between the consecutive pulses, the memory storage time, and the time delay in the delay line are all set to be identical. This allows part of the incoming signal, which was not stored due to limited efficiency, to interfere with the first reference pulse, and the retrieved signal to interfere with the second reference pulse. The phase difference between these two pulses, acquired in a single shot within tens of ns, is thus insensitive to interferometer drifts, which occur over a much longer time scale. We then average this value over hundreds of repetitions.



 
\emph{Author contributions--} R.F., O.L., O.D., S.K., and E.P. contributed to the experimental design, construction, data collection and analysis of this experiment. I.C., R.F., O.L., and E.P. developed the theoretical framework supporting the experiment. E.P. and O.F. supervised the entire project. All authors discussed the results and contributed to writing
the manuscript.

\emph{Competing interests}-- The authors declare no competing interests.

\emph{Acknowledgements--} We acknowledge financial support by the Israel Science Foundation and ICORE, the
European Research Council starting investigator grant QPHOTONICS 678674, the Pazy Foundation, the Minerva
Foundation with funding from the Federal German Ministry
for Education and Research, and the Laboratory in
Memory of Leon and Blacky Broder. I.C. acknowledges support from Marie Skodowska-Curie grant agreement no. 785902.

\newpage
\bibliographystyle{latexmkrc}
\bibliography{bib_protection_arxiv.bib}

\renewcommand\theequation{S\arabic{equation}}
\setcounter{equation}{0}  
\renewcommand\thefigure{S\arabic{figure}}
\setcounter{figure}{0}  
\renewcommand\thetable{S\arabic{table}}
\setcounter{table}{0}  
\renewcommand{\thesection}{S-\Roman{section}}
\setcounter{section}{0}
\renewcommand\thesubsection{\thesection.\arabic{subsection}}
\renewcommand\thesubsubsection{\thesubsection(\alph{subsubsection})}
\clearpage

\onecolumngrid
\widetext

\section*{Supplementary Information: Continuous protection of a collective state from inhomogeneous dephasing  }
\section*{Derivation of protection conditions and sensitivity analysis}
Let $\{\down,\up,\sens\}$ be a three-level system, comprising a qubit $\{\down,\up\}$ and a sensor state $\sens$. Let the frequencies of the transitions $\down-\up$ and $\down-\sens$ be inhomogeneously shifted, with the magnitude of the shifts linearly dependent on the same inhomogeneous variable $\delta$. We assume (without loss of generality, see footnote \footnote{ Consider three states $\mid\psi_n\rangle$ ($n=1,2,3$) with energies  $\hslash \omega_n(\delta^*)=\hslash(\omega_n^0+a_n\delta^*)$, arranged such that $a_1\le a_2 \le a_3$. If $a_1=a_2$ or $a_2=a_3$, then the corresponding transition is already protected from dephasing due to changes in $\delta^*$. Otherwise, denote $n_\pm=2\pm\mathrm{sign}(a_1-2a_2+a_3)$ and define $\down\equiv\mid\psi_2\rangle$, $\up\equiv\mid\psi_{n_-}\rangle$, $\sens\equiv\mid\psi_{n_+}\rangle$, $\delta\equiv(a_2-a_{n_-})\delta^*$, and $s\equiv(a_{n_+}-a_2)/(a_2-a_{n_-})$.}) that $\down-\up$ and $\down-\sens$ shift to opposite directions and that $\down-\sens$ is at least as sensitive as $\down-\up$ to the inhomogeneity. Hereafter, we take the ground state $\down$ as the frequency reference and examine the transition frequencies relative to it. In a frame rotating with the unshifted frequencies, the Hamiltonian in the subspace $\{\up,\sens\}$ is given for any constituent in the ensemble as ($\hslash$=1)
\beq
H_0= \delta \bla{\begin{array}{ccc}
-1 & 0 \\
0  & s
\end{array}}.
\label{z_noise}
\eeq 
Here $\delta$ is a random (inhomogeneous) variable with a standard deviation $\sigma$, and the sensitivity parameter $s\ge1$ is defined by construction. We aim to add coupling fields in this subspace that will render (at least one) protected state, \emph{i.e.}, a dressed state with a reduced sensitivity to $\delta$. Note that we do not explicitly use $\sigma$ in the following; rather, it is implicitly understood that $\sigma$ should substitute for $\delta$ when assessing the validity of various expansions in leading orders in $\delta$.

\subsection{Protection by a single-tone dressing field}
A single classical field that dresses the transition $\up-\sens$ with a Rabi frequency $\Omega$ and detuning $\Delta$ yields the Hamiltonian
\beq
H=H_0+ 
\bla{\begin{array}{ccc}
\Delta & \Omega \\
 \Omega & 0
\end{array}  }.
\eeq
By diagonalizing $H$, we obtain the new transition frequencies
\beq
\omega_{\pm}=\frac{1}{2}\blb{\Delta+\delta\bla{s-1}\pm\sqrt{\blb{\Delta+\bla{s+1}\delta}^2 +4\Omega^2}}.
\eeq
By requiring a vanishing derivative with respect to $\delta$ of one of these frequencies $(d\omega/d\delta)|_{\delta\rightarrow 0} = 0$, we obtain the protection condition [Eq.~(1) in the main text]
\beq
\bla{\frac{\Omega}{\Delta}}^2=\frac{{s}}{\bla{s-1}^2}.
\label{eq:condSingle}
\eeq 
In the case of a sensor state with high sensitivity $s\gg 1$, this condition reduces to $(\Omega/\Delta)^2=1/{s}$. 

\begin{figure}[bt!]
    \centering
    \includegraphics[width=0.9\textwidth]{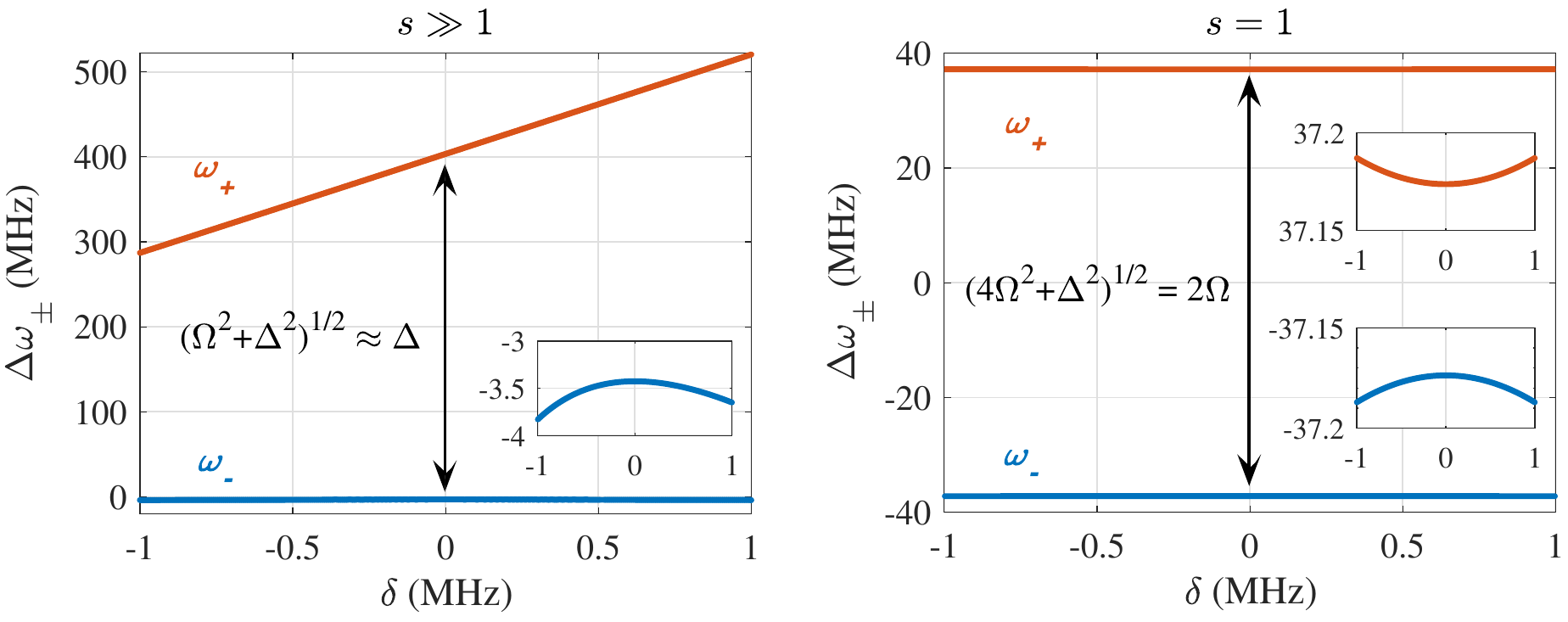}
    \caption{Dressed-state spectrum for a single-dressing protection. Transition frequency shift as a function of the inhomogeneous shift $\delta$, after introducing a dressing field with Rabi frequency $\Omega=37$ MHz and detuning $\Delta$. Left panel: For high sensor sensitivity  ($s=110\gg 1$), there exists one protected state (insensitive to $\delta$ to first order). The energy gap is dominated by the detuning $\Delta=400$ MHz, which is much larger than $\Omega$. Right panel: For $s=1$, there exist two protected states. Optimal protection is achieved with a resonant dressing field ($\Delta=0$), and thus the energy gap is dominated by $\Omega$. Insets magnify the spectrum near $\delta=0$, revealing the magnitude of the second-order shift.}
    \label{fig:single_dressing}
\end{figure}

An example for the spectrum of the two dressed states under the protection condition is shown in Fig. \ref{fig:single_dressing}. For the general case $s \neq 1$, the two states $\ket{\uparrow},\ket{S}$ are not equally mixed, and we obtain a single protected state. Conversely, when $s=1$, optimal protection is reached with a resonant dressing $\Delta=0$ and with the two states equally mixed, and therefore we obtain two protected states, which can be employed to form a qutrit. In this regime, our scheme is akin to the traditional two-level continuous dynamical decoupling (CDD)  \cite{Lidar2004pra}.

While protecting up to first order in $\delta$, the single dressing field adds a qubit transition frequency shift of second-order in the inhomogeneous shift $\delta$ 
\beq
\Delta \omega_{\delta^2}=\frac{s\delta^2}{\Omega}\frac{\sqrt{s}}{s+1},
\label{second_ord}
\eeq
which is inversely proportional to the available $\Omega$. In addition, the protection is subjected to drive noise in the form of amplitude fluctuations $\Omega \rightarrow \Omega+\delta_\Omega$, giving rise to a transition frequency shift
\beq 
\Delta \omega_{\delta_\Omega}=2\delta_\Omega\frac{\sqrt{s}}{s+1}.
\label{Rabi1}
\eeq

For high sensitivity $s\gg 1$, the drive noise is attenuated by $1/\sqrt{s}$, while the second-order inhomogeneous shift is enhanced by $\sqrt{s}$. These scalings are summarized in table \ref{tab:sens} (left section) and compared to the residual transition frequency shifts under the other protection schemes, presented in the following sections.

\begin{table}[h!]
\caption{Higher-order corrections for the qubit transition frequency shifts: first-order in the drive noise ( O($\delta_{\Omega}$)), second- and third- order in the inhomogeneous shift (O($\delta^2$), O($\delta^3$)) and their cross-term (O($\delta\delta_\Omega$)).
\\}
\begin{tabular}{ |c||c|c||c|c|c|c||c|c|c|c|  }
\colrule
 & \multicolumn{2}{c||}{single-tone dressing} & \multicolumn{4}{c||}{two-tone dressing} & \multicolumn{4}{c|}{stepwise phase} \\ [0.5ex]

& O($\delta_\Omega$) &  O($\delta^2$) & O($\delta_\Omega$) &  O($\delta^2$) & O($\delta \delta_\Omega$) &  O($\delta^3$) & O($\delta_\Omega$) &  O($\delta^2$) & O($\delta \delta_\Omega$) &  O($\delta^3$) \\ [0.5ex]
\colrule
& & & & & & & & & &\\ [-2ex]
 $s=1$	& $\delta_\Omega$ & $\frac{1}{2}\frac{\delta^2}{\Omega}$ & 0 & 0
 & $1.25\frac{\delta\delta_\Omega}{\Omega}$ & ${(\scriptstyle 0.06\sim0.4)}\frac{\delta^3}{\Omega^2}$ & 0 & 0 & $\frac{\delta\delta_\Omega}{\Omega}$ & $\frac{1}{2}\frac{\delta^3}{\Omega^2}$\\[1ex]

 $s\gg 1$ &	$\frac{2}{\sqrt{s}}\delta_\Omega$ & $\sqrt{s}\frac{\delta^2}{\Omega}$ & 0 & 0	
 & $2\frac{\delta\delta_\Omega}{\Omega}$ & $s\frac{\delta^3}{\Omega^2}$ & 0 & 0 & $2\frac{\delta\delta_\Omega}{\Omega}$ & $\frac{6s}{5}\frac{\delta^3}{\Omega^2}$\\[1ex]
\colrule
 \end{tabular}
 \label{tab:sens}
 \end{table}

\subsection{Protection by a two-tone dressing field}
A dressing field comprising two tones with symmetric detuning $\pm\Delta$ around the $\up-\sens$ transition can protect the $\down-\up$ transition better than a single-tone dressing, that is, up to higher orders in the inhomogeneous shift $\delta$ and in the drive noise $\delta_\Omega$. With this field, the time-dependent Hamiltonian in the rotating frame is given by $H(t)=H_0+H_2(t)$, where 
\beq
H_2(t)= 
\sqrt{2}\Omega \cos(\Delta t+\phi)\bla{\begin{array}{ccc}
 0& 1\\
 1 & 0
\end{array}  },
\label{Hd2}
\eeq
where $\phi$ is the phase between the two tones of the dressing at $t=0$. The $\sqrt{2}$ factor appears above in order to maintain the overall dressing power as in the single-dressing scheme. 

\subsubsection{Solution for $s\gg 1$}
Before presenting the general solution, we consider the limit $s\gg 1$. Since $\Delta \gg\Omega$, we can solve the time-dependent Hamiltonian using the effective Hamiltonian technique~\cite{EH}, where each field gives rise to a standard light shift,
\beq
H_{\pm\Delta}=\frac{\Omega^2/2}{\pm\Delta+\delta (s+1)} \bla{\begin{array}{ccc}
 -1&0 \\
  0& 1
\end{array}  } 
\approx \frac{\Omega^2}{\pm 2\Delta}\bla{1-\frac{s\delta}{\pm\Delta}+\blb{\frac{s\delta}{\Delta}}^2}\bla{\begin{array}{ccc}
 -1&0 \\
  0& 1
\end{array}}.
\eeq
When summing both contributions $H\approx H_{+\Delta}+H_{-\Delta}$, we are left with only the odd orders in $\delta$. The first order compensates for the inhomogeneous shift terms in the bare Hamiltonian $H_0$ when the protection condition
\beq
\frac{\Omega^2}{\Delta^2}=\frac{1}{s}
\label{double_light_shift}
\eeq 
is satisfied. This condition coincides with that of the single-dressing scheme [Eq.~(\ref{eq:condSingle})] for $s \gg 1$. However, unlike for single dressing, here the residual inhomogeneity is of third order in $\delta$, as both the first and the second orders are elimniated. In addition, since the zeroth order of the light shift is eliminated as well, the double dressing scheme is also protected from the first-order contribution of the drive noise $\delta_\Omega$ (the fluctuations in $\Omega$). A general solution and a complete sensitivity analysis for the general case are presented below.

\subsubsection{Solution for general $s$ using the Magnus expansion}
We use the Magnus expansion to calculate both the exact protection condition for a general $s$ and the remaining high-order noise contributions. We start by writing the bare Hamiltonian \eqref{z_noise} as
\beq
H_0=\frac{\delta}{2}\bla{ \blb{s-1} \mathbb{I} -\blb{s+1}\sigma_z },
\label{rewrite}
\eeq
where $\sigma_z$ is the Pauli-$z$ operator, and $\mathbb{I}$ is the identity. 
To ease the derivation, we rotate the system around the $y$ axis, such that $\sigma_x \rightarrow S_z$ and $\sigma_z \rightarrow -S_x$. The rotated Hamiltonian is
\beq
H_y=U_y\bla{H_0+H_2}U_y^\dagger = \frac{\delta}{2}\blb{ \bla{s-1} \mathbb{I} +\bla{s+1}S_x } +\sqrt{2}\Omega S_z \cos \bla{\Delta t +\phi},
\label{Hy12}
\eeq
Next, we move to a time-dependent rotating frame, utilizing $H_2(t)$ to construct the unitary rotation operator 
\beq
U(t)=\exp (-\frac{i}{\hslash}\int_0^{t} U_y H_2(t) U_y^\dagger dt)=e^{-i S_z \sqrt{2}(\Omega/\Delta) [\sin \bla{\Delta t +\phi}-\sin\phi]}
\eeq
and obtain the Hamiltonian in the interaction picture
\beq
H'_y=\frac{\delta}{2}\blb{ \bla{s-1} \mathbb{I} +\bla{s+1}\bla{S_+ e^{i z \blb{\sin \bla{\Delta t +\phi}-\sin\phi}} +\mathrm{H.c.} }} \label{H_interact}
\eeq
with $z={2\sqrt{2}\Omega}/{\Delta}$.  

Using the  Jacobi–Anger expansion 
\beq
e^{\pm i z  \sin\bla{\Delta t +\phi}}=\sum_{n=-\infty}^{\infty}  J_n(z) e^{\pm i n  \bla{\Delta t +\phi}},
\label{Jacobi-Anger}
\eeq
we expand Eq.~\eqref{H_interact} in leading orders of the Magnus expansion. The first order reads
\beq
H^{(1)}=\frac{1}{t}\int_0^t H'_y(t_1)dt_1=\frac{\delta}{2}\blb{ \bla{s-1} \mathbb{I} +J_0(z)\bla{s+1}\bla{S_+ e^{-iz\sin\phi}+S_- e^{iz\sin\phi}} }.
\label{dressed}
\eeq
Since $\bla{S_+ e^{-iz\sin\phi}+S_- e^{iz\sin\phi}} $ is a sum of Pauli matrices, it has two eigenvalues $\pm1$. To find the protection condition [Eq.~(2) in the main text], we require that the eigenvalues of $H^{(1)}$ be independent of $\delta$, which occurs when
\beq
J_0(z)=\frac{s-1}{s+1}.
\label{condition}
\eeq
For $s\gg 1$, the condition \eqref{condition} can be expanded to the leading order in $z$: 
\beq
1-\frac{z^2}{4}+O(z^4)=1-\frac{2}{s} + O(s^{-2}),
\eeq
where we recover the previous (light shift) condition \eqref{double_light_shift}. For $s=1$, as before, the first-order dependence on $\delta$ vanishes for \emph{both} dressed states, and we obtain two protected states. Therefore, again, the whole three-level system is protected \cite{Sutherland2019,Srinivas2019}. 

For a general $s\neq 1$, the protected dressed state is rotated by $z\sin\phi$ around the $x$ axis in the $\{\up,\sens\}$ subspace, 
\beq
\ket{P}=\exp\bla{-i \sigma_x \frac{z\sin\phi}{2}}\up.
\eeq
When the fields are in-phase ($\phi=0$), or for high sensitivity $s \gg 1 $ (as in our experiment), the dressed state is approximately the bare state $\up$. Otherwise, in order to efficiently couple to the protected dressed state, the phase term $(z \sin \phi)/2$ needs to be well defined, and it would determine the adiabatic condition for switching on and off of the protection fields \cite{Roos}. 


\subsubsection{Higher-order transition frequency shift contributions}
To find the higher-order noise terms, it is beneficial to make the distinction between noise terms that are parallel and perpendicular to the dressed-state basis. For brevity, we define the pauli-$x$ operator in the $z\sin\phi$-dependent (dressed state) basis as $F_x=\bla{S_+ e^{-iz\sin\phi}+S_- e^{iz\sin\phi}}$ an rewrite Eq.~\eqref{dressed} as
\beq
H^{(1)}=\frac{\delta\bla{s-1}}{2}\bla{\mathbb{I} +F_x}.
\label{dressed2}
\eeq
The protected dressed state is the eigenstate of $F_x$ with the eigenvalue $-1$. The perpendicular pauli operators are $F_y=-i\bla{S_+ e^{-iz\sin\phi} - S_- e^{iz\sin\phi}}$, and $F_z=S_z$.

We begin with the cross term of the inhomogeneous shift $\delta$ and the drive amplitude fluctuations $\delta_\Omega$,
\beq
H^{(1)}_{\delta\cdot\delta_\Omega}=[-J_1(z)F_x +\sin\phi J_0(z)F_y]\frac{2\sqrt{2}}{\Delta}\frac{s+1}{2}  \delta\delta_\Omega.
\label{Rabi_Doppler}
\eeq
With respect to Eq.~\eqref{dressed2}, this noise term has both a parallel $\bla{\propto F_x}$ and a perpendicular $\bla{\propto F_y}$ contributions. Notably, the perpendicular contribution vanishes when $\phi\rightarrow 0$. For a general $\phi$, we examine the limits $s=1$ and $s\gg 1$:
\begin{itemize}
\item For $s=1$, according to condition (\ref{condition}), we set $J_0(z)=(s-1)/(s+1)=0$ and are left with only the parallel term
\beq
\Delta \omega_{\delta\cdot\delta_\Omega}\approx 1.25\frac{\delta\delta_\Omega}{\Omega}.
\eeq
\item For $s\gg 1$, we approximate the perpendicular term as $H^{(1)}_{\delta\cdot\delta_\Omega,\perp}\approx \sin\phi \sqrt{2 s} \delta (\delta_\Omega/\Omega) F_y $.
From Eq.~\eqref{dressed2}, if the energy gap $\Delta \omega= \delta\bla{s-1}$ in the $F_x$ direction is larger than the perpendicular $F_y$ noise, and assuming $\sin\phi\sqrt{2/s}\bla{\delta_\Omega/\Omega}  \ll 1 $, then the perpendicular noise manifests only as a small perturbation $\Delta \omega_{\delta\cdot\delta_\Omega,\perp} \propto \delta (\delta_\Omega/\Omega)^2$. 
Therefore, the leading contribution is the parallel term
\beq
\Delta \omega_{\delta\cdot\delta_\Omega}= 2\frac{\delta\delta_\Omega}{\Omega}.
\eeq
\end{itemize}

We now turn to evaluate the residual second-order contribution of the inhomogeneous shift. We take the next order in the Magnus expansion using the Jacobi-Anger expansion \eqref{Jacobi-Anger}
\beq
H^{(2)}=-\frac{i}{2t}\int_0^t dt_1\int_0^{t_1} dt_2 \blb{H'_y(t_1),H'_y(t_2)}=-2\blb{\frac{\delta\bla{s+1}}{2}}^2  J_0(z)\sum_{n\neq 0} \frac{ J_{n}(z)\cos n\phi}{n \Delta} F_z
\eeq
and again examine different $s$ regimes:
\begin{itemize}
\item For $s = 1$, we have $ J_0(z)=0$, and therefore $H^{(2)}=0$ for any $\phi$.   
\item For general $s > 1$, $H^{(2)}$ contains a perpendicular $\bla{F_z}$ contribution with respect to Eq.~\eqref{dressed2}, which vanishes for $\phi\rightarrow \pi/2$. For a general $\phi$, this perpendicular term is not vanishing. 
\item In the limit $s\gg 1$ and $z/2=\sqrt{2/s} \ll 1$, we expand $H^{(2)}$ to  leading orders  
\beq
H^{(2)} \approx -\blb{\frac{\delta\bla{s+1}}{2}}^2  \frac{2z \cos \phi}{\Delta} F_z \approx  - \frac{\delta^2 s\sqrt{2s} \cos \phi }{\Delta } F_z \approx  - \frac{\delta^2 s\sqrt{2} \cos \phi }{\Omega } F_z.
\label{second_general}
\eeq
Once again, if the energy gap of $H^{(1)}$ [Eq.~\eqref{dressed2}] is larger than this perpendicular term and assuming $\sqrt{2} \cos \phi \delta\ll \Omega  $, we get only a small, third-order correction
\beq
H^{(2)} \approx 2\frac{s\delta^3 \cos^2 \phi  }{\Omega^2}F_x.
\label{sec_AC}
\eeq
\end{itemize}
We find for both $s \gg 1$ and $s=1$ that the double-dressing scheme eliminates the first- and second- order contributions of the inhomogeneous shift. Note that when $s\ne 1$ but not large and $\phi \neq \pi/2$, there will be a non-vanishing second order contribution.

To fully evaluate the contribution of the inhomogeneous shift to third order, we take the third order of the Magnus expansion
\beq
H^{(3)}=-\frac{1}{6t}\int_0^t dt_1\int_0^{t_1} dt_2\int_0^{t_2} dt_3 \bla{\blb{H'_y(t_1),\blb{H'_y(t_2),H'_y(t_3)}} +\blb{H'_y(t_3),\blb{H'_y(t_2),H'_y(t_1)}}     }.
\eeq
For a general $s$, we obtain
\bea
H^{(3)}&=&\blb{\frac{\delta\bla{s+1}}{2}}^3\sum_{n>0} \bld{\frac{4 (J_{2n}(z))^3}{\bla{2n\Delta}^2}\cos 2n\phi F_x  -  \frac{4 (J_{2n-1}(z))^3}{\bla{(2n-1)\Delta}^2}\sin (2n-1)\phi F_y} \nonumber\\ 
&&-\blb{\frac{\delta\bla{s+1}}{2}}^3J_0(z)\sum_{n>0} \frac{2(2+\cos 2n\phi)  (J_{n}(z))^2}{\bla{n\Delta}^2}F_x\\
&&+\blb{\frac{\delta\bla{s+1}}{2}}^3(J_0(z))^2\sum_{n>0} \frac{8J_{2n-1}(z)}{\bla{(2n-1)\Delta}^2}   \sin (2n-1)\phi F_y \nonumber\\
&&-\blb{\frac{\delta\bla{s+1}}{2}}^3J_0(z)\sum_{n\neq m \neq 0}\frac{4 J_n(z)J_m(z)}{\Delta^2} \bld{\frac{\cos n\phi\cos m\phi}{nm} F_x +\frac{m\cos n\phi\sin m\phi-n \sin n\phi\cos m\phi}{nm(m-n)} F_y} .\nonumber
\eea
For $s = 1$ this becomes
\beq
H^{(3)}=\delta^3\sum_{n>0} \bld{\frac{4 (J_{2n}(z))^3}{\bla{2n\Delta}^2}\cos 2n\phi F_x-\frac{4 (J_{2n-1}(z))^3}{\bla{(2n-1)\Delta}^2}\sin (2n-1)\phi F_y} \approx \frac{\delta^3}{\Omega^2} \bla{0.06 \cos2\phi F_x - 0.4\sin\phi F_y}.
\eeq
Where we have used the fact that around the protection condition the sum is well approximated by the first term. Since for $s=1$ the energy gap from Eq.~\eqref{dressed2} is vanishing, both terms ($F_x$ and $F_y$) contribute to the noise. The correction to the transition frequency due to inhomogeneous shift thus varies between 
\beq
\Delta \omega_\delta \in (0.06,0.4) \cdot \frac{\delta^3 }{\Omega^2},
\eeq
depending on the phase $\phi$ . The lower noise amplitude is obtained when the two tones of the dressing are in-phase $\phi=0$.

For $s\gg 1$, the third-order Magnus term becomes
\beq
H^{(3)}\approx \frac{\delta^3 }{\Omega^2} \blb{ -s(2+\cos 2\phi) F_x   +\sqrt{2}\sin\phi s^{3/2} F_y},
\eeq
which should be added to the second-order Magnus term [Eq.~\eqref{sec_AC}]
\beq
H^{(2)}_{\mathrm{eff}}+H^{(3)}\approx \frac{\delta^3 }{\Omega^2} \bla{ -s F_x   +\sqrt{2}\sin\phi s^{3/2} F_y}.
\eeq
Once again, due to the large energy gap in the $F_x$ direction, the perpendicular term ($\propto F_y$) produces a negligible perturbation. Therefore when $s \gg 1$, the first non-vanishing order correction to the transition frequency due to the inhomogeneous shift is
\beq
\Delta \omega_{\delta} \approx \frac{s \delta^3 }{\Omega^2}.
\eeq
A summary of these leading terms is presented in Table \ref{tab:sens} (middle section).
The decay of coherence due to this third-order correction in non-Gaussian, due to the non-linear coupling to shift $\delta$, and can be found by taking the Fourier transform of the transition frequency spectrum. 
%


\subsection{Protection by a dressing field with a stepwise-modulated phase}
In our experiment, we did not explicitly employ the two-tone (symmetrically detuned) dressing of Eq.~\eqref{Hd2}. Instead, we passed a single-tone, resonant ($\Delta=0$) field through an electro-optic modulator and alternated its phase stepwise between 0 and $\pi$ with a half-period $T/2$. This forms a multi-tone signal with a vanishing carrier, dominant first-order side bands, and additional higher-order side bands. The  power spectral density of this signal thus highly resembles the pure two-tone case discussed in the previous section. However, an exact solution may be obtained for this case. The Hamiltonian alternates between $H_+$ and $H_-$, where $H_\pm=H_0\pm\Omega \sigma_x$. In analogy with the phase $\phi$ of the two-tone dressing in the previous section, here we control the duration of the initial interval (relative to the qubit initialization time $t=0$). We can describe the unitary evolution over the first full period $T$ with
\beq
U_{T}=e^{-i H_{-} [1-f(\phi)] T/2} e^{-i H_{+} T/2} e^{-i H_{-}  f(\phi)T/2},
\eeq  
where $f(\phi)=\cos^2\bla{\pi/4-\phi/2}$. For an evolution time $t=n T$, the evolution operator is $U(t)=\bla{U_{T}}^n$. One can see that the stepwise dressing with period $T$ corresponds to the two-tone dressing with detuning $\Delta=2\pi/T$.
To expand $U_{T}$ in leading orders of the inhomogeneous shift $\delta$ and the drive noise $\delta_{\Omega}$, we define an effective Hamiltonian $H_{eff}$ by  $U_{T}=\exp\bla{-i H_{\mathrm{eff}} 2\pi/\Delta }$, such that $H_{\mathrm{eff}}$ can be expanded in orders of $\delta$. The first order reads 
\beq
H^{(1)}={\delta}\frac{s-1}{2} \mathbb{I} - {\delta}\frac{s+1}{2}\frac{\sin\bla{\pi\Omega/\Delta }}{\pi\Omega/\Delta} \bld{\cos\blb{(\pi\Omega/\Delta)\sin\phi}\sigma_z-\sin\blb{(\pi\Omega/\Delta) \sin\phi}\sigma_y}.
\label{eq:H1stepwise}
\eeq
Therefore, for protection up to first order we require: 
\beq
\sinc \bla{\pi\Omega/\Delta}=\frac{s-1}{s+1},
\label{condition_p}
\eeq 
where $\sinc(x)=\sin(x)/x$. The protected dressed state is the eigenstate of the operator $F_x=-\cos\blb{(\pi\Omega/\Delta)\sin\phi}\sigma_z-\sin\blb{(\pi\Omega/\Delta) \sin\phi}\sigma_y$ with the eigenvalue $-1$ [Analogous to Eq.~\eqref{dressed2}]. 
For $s\gg 1$, the condition \eqref{condition_p} can be expanded to leading order,
\beq
1-\frac{(\pi\Omega/\Delta)^2}{6}+O((\pi\Omega/\Delta)^4)=1-\frac{2}{s} + O(s^{-2}).
\eeq
We thus arrive at the condition $(\Omega/\Delta)^2 = {12/\pi^2 s} \approx 1/{s}$, in agreement with both the one-tone and two-tone dressings at the $s\gg 1$ limit.

\subsubsection{Higher-order noise contributions}
We maintain the notations of parallel ($\propto F_x$) and perpendicular ($\propto F_y,F_z$) noise terms in the dressed-state basis, where now $F_y=\cos\blb{(\pi\Omega/\Delta) \sin\phi}\sigma_y-\sin\blb{(\pi\Omega/\Delta) \sin\phi}\sigma_z$ and $F_z=\sigma_x$. 
The cross term of the drive noise together with the inhomogeneous shift is
\beq
H^{(1)}_{\delta\cdot\delta_{\Omega}}= \frac{1}{\Omega}\bld{\bla{\cos(\pi\Omega/\Delta) -\sinc (\pi\Omega/\Delta) }F_x - \sin\phi \sin (\pi\Omega/\Delta)  F_y    } \cdot\blb{\frac{s+1}{2}\delta\delta_\Omega}
\label{Rabi_Doppler_pulse}
\eeq 
As in the two-tone case, the perpendicular contribution vanishes when $\phi\rightarrow 0$. For a general $\phi$, we examine the limits $s=1$ and $s\gg 1$:
\begin{itemize}
\item For $s=1$, under the protection condition \eqref{condition_p}, the perpendicular term vanishes and we are left with the parallel term
\beq
\Delta \omega_{\delta\cdot\delta_\Omega}= \frac{\delta\delta_\Omega}{\Omega}.
\eeq

 \item For $s\gg 1$, this perpendicular term gives rise to a higher-order perturbation when the energy gap $\delta(s+1)$ [Eq.~\eqref{eq:H1stepwise}] is large enough. The leading contribution in the parallel term is:
 \beq
\Delta \omega_{\delta\cdot\delta_\Omega}= 2\frac{\delta\delta_\Omega}{\Omega}.
\eeq
\end{itemize}

To second order in $\delta$, the effective Hamiltonian reads
\beq
H^{(2)}= \frac{1}{\Omega} \sinc \bla{\pi\Omega/\Delta}\bla{\cos \pi\Omega/\Delta - \cos\blb{(\pi\Omega/\Delta)\sin\phi} }\cdot \bla{\frac{s+1}{2}{\delta}}^2 F_z.
\label{second_general_p}
\eeq
As in the two-tone case, the second-order contribution vanishes for $\phi\rightarrow \pi/2$ or for $s=1$. 
For $s\gg 1$, we are left with \mbox{$H^{(2)}\approx -\cos^2\phi (3/2)(\delta^2/\Omega) F_z$}, which is perpendicular to the dressed state basis and thus only contributes a third-order correction 
\beq
H^{(2)}\approx  \cos^4\phi\frac{9s}{4\Omega^2}\delta^3 F_x,
\eeq
assuming a large enough energy gap $\delta(s+1)$. For a general $\phi$ and $s$, the second-order term does not vanish.
The effective Hamiltonian to the third order, for $s=1$, is $H^{(3)}=-\delta^3/(2\Omega^2) F_x$. For $s\gg 1$ the third order term is
\beq
H^{(3)}=-3s \frac{64 +120 \cos^4 \phi}{160\Omega^2} \delta^3 F_x 
\eeq
and, together with the higher order contribution of $H^{(2)}$, we obtain $H^{(2)}+H^{(3)}\approx - \frac{6s}{5\Omega^2} \delta^3 F_x $. the first non-vanishing order correction to the transition frequency due to the inhomogeneous shift is thus
\beq
\Delta \omega_\delta=\frac{6s}{5\Omega^2} \delta^3
\eeq
A summary of these leading terms is presented in Table \ref{tab:sens} (right section).
\section*{Fig. S2: Supplementary information for $T_{inhom}$ evaluation and calibration measurements}
\begin{figure}[th!]
    \centering
    \includegraphics[scale=0.7]{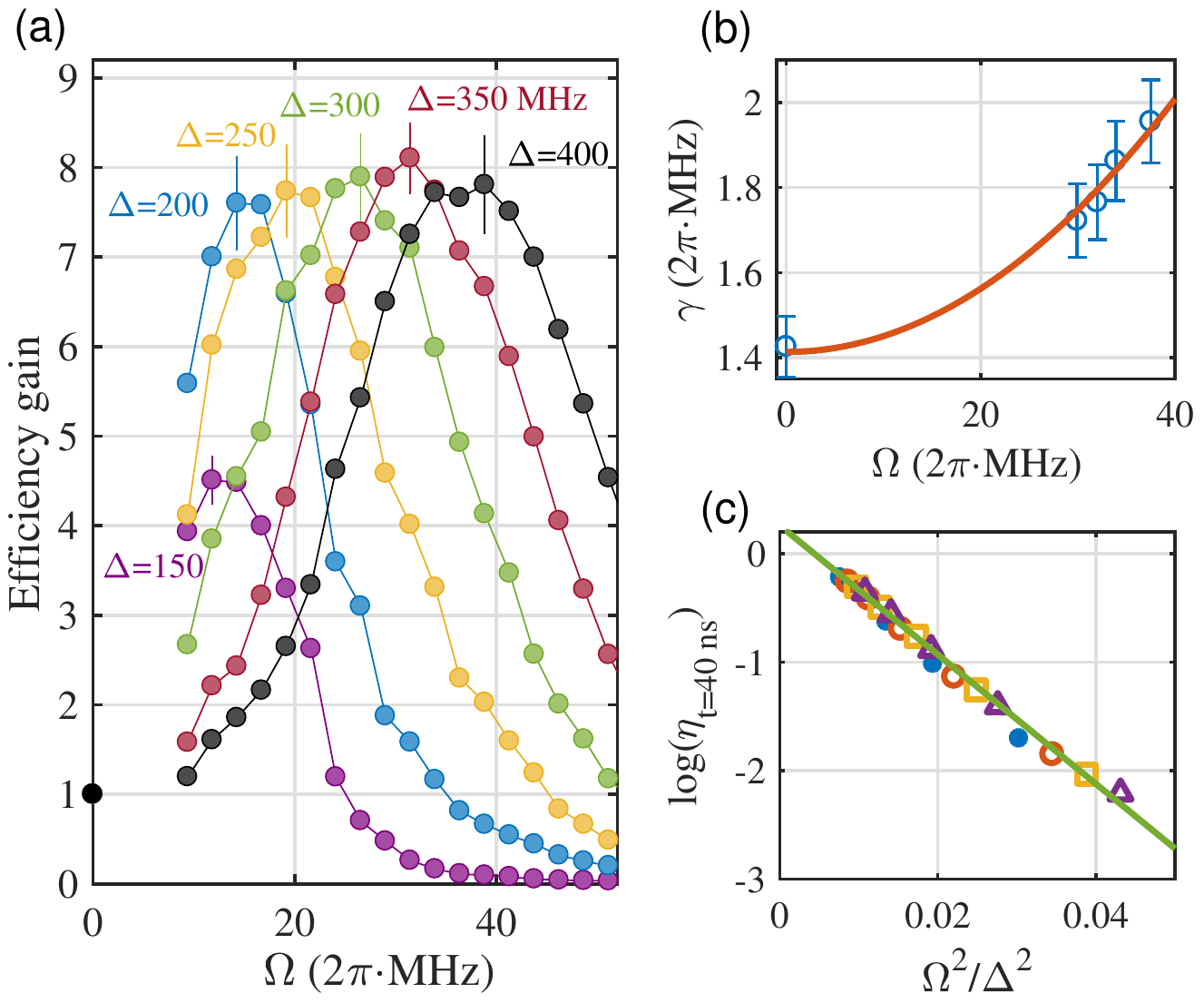}
    \caption{\textbf{Efficiency enhancement and loss with a dressing field at intermediate detunings.} (a) Gain in efficiency at a long storage time ($t=250$ ns) with double dressing protection as a function of dressing Rabi frequency $\Omega$. Solid lines are guide to the eye. The gain saturates due to a slight increase in homogeneous decoherence at large dressing powers. (b) Added homogeneous decoherence rate $\gamma$ for several dressing Rabi frequencies $\Omega$ at detuning $\Delta=2\pi\cdot350$ MHz. Solid line is a fit to a quadratic model. (c) Efficiency loss at a short storage time ($t=40$ ns) for different powers (marked as different markers) and for several detunings. All data sets collapse when plottted against $\Omega^2/\Delta^2$ and fit well to the scattering model model $\gamma=\gamma_0+\Gamma\Omega^2/\Delta^2$ across a wide range of the parameter $\Omega/\Delta$ }
    \label{fig:my_label}
\end{figure}

\providecommand{\url}[1]{\texttt{#1}}
\providecommand{\urlprefix}{URL }
\providecommand{\eprint}[2][]{\url{#2}}










\end{document}